# Stop_times based Routing Protocol for VANET

Hafez Moawad
Faculty of Computer and Information science Ain shams University
Cairo, Egypt

Eman Shaaban
Faculty of Computer and Information science Ain shams University
Cairo, Egypt

Zaki Taha Fayed
Faculty of Computer and Information science Ain shams University
Cairo, Egypt

## ABSTRACT
Vehicular Ad hoc Network (VANET) is a special class of Mobile Ad hoc Network (MANET) where vehicles are considered as MANET nodes with wireless links. The key difference of VANET and MANET is the special mobility pattern and rapidly changeable topology. There has been significant interest in improving safety and traffic efficiency using VANET. The design of routing protocols in VANET is important and necessary issue for support the smart ITS. Existing routing protocols of MANET are not suitable for VANET. AOMDV is the most important on demand multipath routing protocol. This paper proposes SSD-AOMDV as VANET routing protocol. SSD-AOMDV improves AOMDV to suit VANET characteristics. SSD-AOMDV adds the mobility parameters: Stop_times, Speed and Direction to hop count as new AOMDV routing metric to select next hop during the route discovery phase. Stop_times metric is added to simulate buses mobility pattern and traffic lights at intersections. Simulation results show that SSD-AOMDV achieves better performance compared to AOMDV.

## General Terms
Wireless Ad hoc networking

## 1. INTRODUCTION
Nowadays, the safety of motor vehicle has been paid more and more attention by the whole society. The increasing problem of accident and traffic jam necessitates the adoption of Intelligent Transportation Systems (ITS). A Vehicular Ad-Hoc network is a form of Mobile ad-hoc Networks MANETs, to provide communication among nearby vehicles and between vehicles and nearby fixed equipment i.e. roadside equipment as in Fig. 1. In a VANET, the vehicles are considered as nodes. Vehicle velocities are also restricted according to speed limits, level of congestion in roads, and traffic control mechanisms (e.g., stop signs and traffic lights). Future vehicles can be equipped with devices have longer transmission ranges. Rechargeable source of energy, extensive on-board storage capacities and processing power are not issues in VANET as they are in MANET. The main goal of VANET is providing safety and comfort for passengers. Besides safety applications VANET also provide comfort applications to the road users. For example, weather information, mobile e-commerce, Internet access and other multimedia applications. The vehicles of a VANET are equipped with the DSRC (Dedicated Short Range Communication). Vehicles can move along the same road way and transmit information or receive ¬¬¬¬-information. Each vehicle equipped with VANET device will be a node in the Ad-hoc network and can receive & relay other messages through the wireless network. VANET is one of the influencing areas for the improvement of ITS in order to provide safety and comfort to the road users. Collision warning and in place traffic view will give the driver essential tool to decide the best path along the way.

MANET and VANET are characterized by the movement and self organization of nodes. The key difference of VANET and MANET is the special mobility pattern and rapidly changeable topology of VANET. Also, MANET nodes cannot recharge their battery power where VANET has no power constraint for nodes.

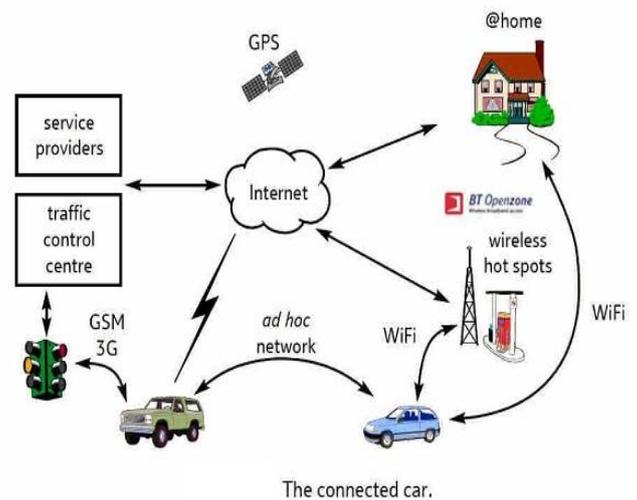

**Fig. 1: VANET communication**

The design of effective vehicular communications poses a series of technical challenges. Guaranteeing a stable and reliable routing mechanism over VANETs is an important step toward the realization of effective vehicular communications. One of the critical issues consists of the design of scalable routing algorithms that are robust to frequent path disruptions caused by vehicles' mobility. Existing routing protocols, which are traditionally designed for MANET, do not make use of the unique characteristics of VANETs and are not suitable for vehicle-to-vehicle communications over VANETs. Topology-based and position-based routing is two strategies of data forwarding commonly adopted for multi-hop wireless networks [1], [2]. Topology-based protocols use the information of available network links for packet transmission. Every node has to maintain the routing table. Position-based protocols assume that every node is aware of the location of itself, the location of neighbouring nodes, and the location of the destination node. With the increasing availability of GPS-equipped vehicles, Position based Protocol is getting more convenient. However, position-based protocols developed for MANETs may not directly be applied to vehicular environments, due to the unique vehicular network characteristics.

One good way of data forwarding in VANET is to modify MANET routing protocols and make it suitable for vehicular environment. There are many routing protocols for ad hoc networks [3], [4], [5]. One of the most well-known is AODV [6], [7], [8], [9]. Ad-hoc On-demand Multipath Distance Vector Routing (AOMDV) protocol is an extension to AODV protocol for computing multiple loop-free and link disjoint paths [10].

This paper proposes SSD-AOMDV as VANET routing protocol. AOMDV is the most important on demand multipath routing protocol. SSD-AOMDV improves AOMDV to suit VANET characteristics. SSD-AOMDV adds the mobility parameters: stop_ times, speed and direction to hop count as new AOMDV routing metric to select next hop during the route discovery phase. Simulation results show that SSD-AOMDV achieves better performance compared to AOMDV.

The remainder of this paper is structured as follows. Section II introduces AOMDV routing protocol. Section III surveys the related researches conducted in enhancing MANET routing protocols for V2V communication. Section IV introduces the proposed scheme SSD-AOMDV. Section V presents the simulation results and discussions. The paper is concluded in Section VI.

## 2. RELATED WORKS

This section surveys the related researches conducted in enhancing MANET routing protocols for V2V communication. In [11], three mobility parameters: position, direction and speed to select the next hop for routing. In that method, direction has the highest priority in selecting next hop during a route discovery phase. With respect to mobility model, if a node has same direction with source and/or destination nodes, it might be selected as a next hop. Position is another parameter that was used for the next hop selection.

S-AOMDV routing protocol is designed to make use of advantages of multi-path routing protocol, such as fault-tolerant and load balance [12]. The routing metric combining hop and speed is proposed with consideration of vehicle driving information employment and delay reduction. Compared with AOMDV, simulation results show that S-AOMDV achieves better performance. Especially with high load (>=8 packet/s), the performance metrics of NRL and Average End-to-End Delay are reduced by 11.1% and 11.9%, respectively.

In R-AOMDV routing protocol proposed in [13], a routing metric combining hop counts and retransmission counts at MAC layer is proposed with consideration of link quality and delay reduction. Based on that routing metric, a cross-layer Ad hoc On-demand Multipath Distance Vector with retransmission counts metric (R-AOMDV) routing protocol is designed to make use of advantages of multi-path routing protocol, such as decrease of route discovery frequency.

SD-AOMDV routing protocol proposed in [14] adds the mobility parameters: speed and direction to hop count as new AOMDV routing metrics to select next hop during the route discovery phase. Simulation results show better performance achieved by SD-AOMDV in general. End-to-end delay decreased by 76.5%. Packet Delay Fraction PDF has been increased by 11.9% compared to AOMDV. However Normalized Routing Load NRL with SD-AOMDV has been increased by 29.4% compared to AOMDV.

## 3. PROPOSED SSD-AOMDV

Proposed SSD-AOMDV improves AOMDV to suit VANET characteristics. SSD-AOMDV adds the mobility parameters: speed, direction and stop_ times to hop count as new AOMDV routing metrics to select next hop during the route discovery phase. Stop_times metric is added to simulate buses mobility pattern and traffic lights at intersections.

To calculate stop_times parameter, it depends on the movement history of vehicles. Some nodes have longer travel time and more of stop_times like buses. Other nodes have less of stop_times like cars.

When a source node requires sending a packet to the destination node, SSD-AOMDV gets direction, speed and stop_times of the source node. Based on direction, speed and stop_times of source, destination and intermediate nodes, paths between source and destination nodes are specified. Because of using Manhattan mobility model, nodes can move in the same direction of source and destination, direction of source, or direction of the destination.

As nodes in VANET move with high speed, their routes are less stable than in MANET. In the other hand, if two nodes move in different directions are communicating together, their communication links break sooner than the state where these nodes move in the same direction. Therefore, if the source and destination are moving in the same direction, the protocol must selects only intermediate nodes that move in the same direction with source and destination. However, if source and destination nodes are moving in a different direction, the protocol must selects only intermediate nodes that move in source direction or destination direction. The protocol also tries to select intermediate nodes that are moving with speed and stop_times close to average of source and destination speed and stop_times. All intermediate nodes have minimum difference between its speed and average speed of source and destination ensuring more path stability. All intermediate nodes have minimum difference between its stop_times and average stop_times of source and destination ensuring more path stability. In the proposed protocol, a route can be selected as forward path between source and destination if all its intermediate nodes move in the same direction with source and/or destination.

SSD-AOMDV combines mobility parameters with hop count as routing metric as follows:

1- For each intermediate node in a disjoint path that moves in the same direction with source and/or destination, the difference between its speed and average speed of source and destination is calculated.

2- For each intermediate node in a disjoint path that moves in the same direction with source and/or destination, the difference between its stop_times and average stop_times of source and destination is calculated.

3- For each disjoint path, speed metric is calculated as the maximum of differences calculated in step 1.

4- For each disjoint path, stop_times metric is calculated as the maximum of the differences calculated in step 2.

5- For all disjoint paths, the forward path is the path with the minimum speed metric. With equal speed metrics values, the path with minimum stop_times metric is selected. With equal speed and stop_times metrics, the path with minimum hop count is selected.

The path satisfies the following condition will be selected to forward packets:

Minimum (Maximum (difference between (Node speed, Average speed of source and destination)[k]), Maximum (difference between (Node stop_times, Average stop_times of source and destination) [k]), hop count).Where K is the number of disjoint paths to the destination node D.

### 3.1 SSD-AOMDV Data Structure

New fields: SrcDir, SrcSpeed, SpeedMetric, and StopMetric are added into original RREQ packet structure specified in AOMDV [15], [11] as shown in table 1.

**Table 1: RREQ packet structure of SSD-AOMDV**

| Source sequence number | SrcDir |
|---|---|
| Hop Count | SrcSpeed |
| SpeedMetric | StopMetric |
| SrcStoptimes | |

Where SrcDir = source direction, SrcSpeed = source speed, SrcStoptimes = source stop_times, SpeedMetric = speed metric with zero initial value, and StopMetric = stop_times metric with zero initial value.

SrcDir, AvgSpeed, AvgStop, SpeedMetric, StopMetric and DestDir fields are added as new fields into original RREP packet structure specified in AOMDV [15], [11] as shown in table 2.

**Table 2: RREP packet structure of SSD-AOMDV**

| Source IP address | AvgStop |
|---|---|
| Destination IP address | SpeedMetric |
| Destination sequence number | StopMetric |
| Hop Count | SrcDir |
| AvgSpeed | DestDir |

Where AvgSpeed = average speed of source and destination, AvgStop = average stop_times of source and destination SrcDir, DestDir = source and destination direction, SpeedMetric = calculated speed metric of destination route, and StopMetric = calculated stop_times metric of destination route.

In the routing table entry, AdvertisedSmetric, AdvertisedStopmetric, DestSpeed, DestStop and DestDir fields are added as new fields into original routing table entry structure specified in AOMDV [15], [11] as shown in table 3.

**Table 3: Routing table entry structure of SSD-AOMDV**

| Dest | DestSpeed |
|---|---|
| Seqno | DestStop |
| Advertised_ Hop Count | AdvertisedStopmetric |
| DestDir | AdvertisedSmetric |
| Route List { list of available paths } | |

Route list has a list of paths for destination SpeedMetric and StopMetric fields are added for each path, as shown in table 4.

**Table 4: Route list entry structure**

| Nexthop | StopMetric |
|---|---|
| HopCountMetric | SpeedMetric |

Where DestSpeed = destination speed, DestStop = destination stop_times, DestDir=destination direction, AdvertisedSmetric=Advertised speed metric and AdvertisedStopmetric = Advertised stop_ times metric.

As a node accepts and maintains multiple routes as obtained by multiple route advertisements, different routes to the same destination may have different HopCountMetric, SpeedMetric and StopMetric. A node must be consistent regarding which one of these multiple metrics is advertised to others. It cannot advertise different HopCountMetric, SpeedMetric or StopMetric to different neighbours with the same destination sequence number. For each destination, only multiple paths that have the same destination sequence number are maintained by a node. With this restriction, a loop freedom invariant similar to AODV is maintained. Once a route advertisement containing a higher destination sequence number is received, all routes corresponding to the older sequence number are discarded.

## 4. SSD-AOMDV DESIGN

SSD-AOMDV is an on-demand routing protocol as AOMDV. When a source node requires a route to a destination, and there are not available paths, the source node will initiate a route discovery process.

### 4.1 Route Discovery Processing

- Source node S broadcasts RREQ routing packet after setting values to new fields as follows:

- SrcDir = current direction of S, SrcSpeed =   current speed of S, SpeedMetric = 0, and StopMetric = 0

- When other nodes receive RREQ packets, they will establish or update reverse paths to the source node S according to SSD-AOMDV routing metric (direction, speed, stop_times  and hops count). These other nodes can be classified into two types: intermediate node I and destination node D.

-If it is an intermediate node I then it establishes a reverse path I~S, searches the routing table for an available forward path I~D to the destination node D. If path l~D exists then node I checks whether it has the same direction of source and/or destination. If TRUE node I discard RREQ packet and sends back RREP packet to S along the reverse path after filling the following new fields:

- SpeedMetric = updated speed metric field of selected forward path in the routing table of I.

- StopMetric = updated stop_times metric field of selected forward path in the routing table of I.

- SrcDir = SrcDir field in RREQ packet.

- DestDir = DestDir field in the routing table entry.

- AvgSpeed = average speed of source speed field in RREQ packet and destination speed in the routing table of I.

- AvgStop = average value of source stop_times in RREQ packet and destination stop_times in the routing table of I.

- If it is an intermediate node and I~D doesn't exist, then node I will rebroadcast RREQ packet after updating SpeedMetric and StopMetric fields of RREQ packet.

- If destination node D receives RREQ packets, it will also establish reverse paths to the source node S. Node D will send RREP packets to node S after filling the new fields in RREP packet as follows:

- SpeedMetric = 0

- StopMetric = 0

- SrcDir = SrcDir field in RREQ packet.

- DestDir = Destination direction.

- AvgSpeed = average speed of source speed field in RREQ packet and speed of node D.

- AvgStop = average stop_times of source stop_times field in RREQ packet and stop_times of node D.

## 4.2 RREP Packet Processing

- If RREP packet is received by an intermediate node, node I checks whether it has the same direction of source and/or destination. If false, node I drops RREP packet, else an RREP packet is forwarded to source after setting SpeedMetric and StopMetric fields of RREP packet as follows:

- SpeedMetric = Max (SpeedMetric of RREP packet, Difference beween(AvgSpeed of RREP packet, speed of the current node)).

- StopMetric = max (StopMetric of RREP packet, Difference between (RREP packet of RREP packet, stop_times of current node)).

- If node I is shared by different link-disjoint paths, and an unused reverse path to node S is available, this reverse path will be selected to forward the RREP packet; otherwise, the RREP packet will be discarded.

- When node S receives RREP packet, SpeedMetric field of RREP packet will record the maximal difference to average speed of source and destination along path D~S. Also StopMetric field of RREP packet will record the maximal difference to average stop_times of source and destination along path D~S. Node S will select a forward path that have minimum SpeedMetric, StopMetric and hop count.

## 5. PERFORMANCE EVALUATIONS

To evaluate the performance of SSD-AOMDV relative to AOMDV and SD-AOMDV, the following performance metrics: end-to-end delay, packet delivery fraction, and normalized routing load are measured against percentage of stopped nodes a. Percentage of stopped nodes is the number of stopped nodes to the total number of nodes. SD-AOMDV routing protocol proposed in [14] adds the mobility parameters: speed and direction to hop count as new AOMDV routing metrics to select next hop during the route discovery phase.

## 5.1 Configuration

The simulation is conducted using NS2.34 [16] and VanetMobisim [17] as a validated vehicular traffic generator. Manhattan is used as Mobility Model. 802.11 is used as MAC layer protocol with transmission range of 250 meters of each node. Traffic pattern consists of 20 CBR/UDP connections between randomly chosen source-destination pairs. a square area of 2000 x 2000 meters for 400 sec simulation time is considered. Speed of vehicles is varying from 10km/h to 90 km/h. Packet generation rate is set to 4 packet/s for Packet size of 512 Bytes. To compare the routing performance with different number of stopped nodes, three scenarios for 60, 70, and 90 nodes are presented. Results are averaged over five simulation runs for each scenario. First the total number of nodes is setting to 60. To simulate buses mobility pattern and traffic lights at intersections, 10%, 20%, 30%, 40%, 50%, 60%, and 70% of the total number of nodes was randomly chosen to stop periodically. Then repeat for 70 and 90 nodes. Finally, the results are averaged for the three scenarios. A snapshot of the mobility is shown in Fig. 2

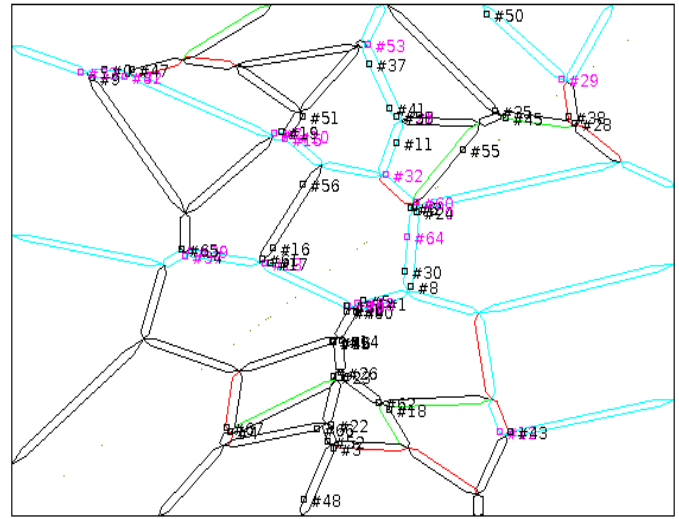

**Fig. 2: A snapshot of the city scenario mobility**

## 5.2 End-to-End delay

End-to-End delay is the average delay in receiving data packets generated by the sources. This includes all possible delays caused by buffering during route discovery, queuing delay at the interface, retransmission delays at the MAC, propagation and transfer times.

Fig. 3 shows average end-to-end delay against the percentage of periodically stopped nodes that simulate buses mobility. Results show that overall average end-to-end delay is improved by 66.9% for SSD-AOMDV and improved by 54% for SD-AOMDV compared to AOMDV. The overall average end-to-end delay is improved by 27% for SSD-AOMDV compared to SD-AOMDV. Since more available valid and stable paths exist in SSD-AOMDV due to considering directions, speed and stop_times in its routing decision, average end-to-end delay is reduced as the percentage of stopped nodes increases. Data packets will be delivered to destinations without route discovery latency. Average end-to-end delay is improved in SD-AOMDV over AOMDV as SD-AOMDV considers speed and direction in its routing decision and generates more stable paths than AOMDV. However when the percentage of periodically stopped nodes less than 20 or higher than 60, there is no significant improvement in average end-to-end delay compared to SD-AOMDV. For less than 20, the probability of stopped nodes existing in the active path is too small to affect its stability. For higher than 20, the probability of stopped nodes existing in the active path is too high to affect its stability as the path is almost stable.

## 5.3 Packet Delivery Fraction

Packet delivery fraction PDF is the ratio of total number of data packets received to the total number of data packets sent by all traffic sources.

Fig. 4 shows that overall average packet delivery fraction is increased by 19.5% for SSD-AOMDV and 15% for SD-AOMDV compared to AOMDV. The overall average packet delivery fraction is improved by 3.8% for SSD-AOMDV compared to SD-AOMDV.

PDF is reduced in SSD-AOMDV since more available valid and stable paths exist in SSD-AOMDV due to considering directions, speed and stop_times in routing decision, and also much more data packets will be delivered to destinations without route discovery latency. However when the percentage of periodically stopped nodes between 20 and 60, percentage of lost packets is decreasing significantly compared to SD-AOMDV. For less than 20 or higher than 60, adding stop_times parameter to SD-AOMDV routing metrics will not affect the stability of the active path significantly.

## 5.4 Normalized routing load

Normalized Routing load NRL is the ratio of total number of routing control packets to the total number of data packet received.

Fig. 5 shows that NRL with SSD-AOMDV has been increased by 30.2 % and increased by 27.5 for SD-AOMDV compared to AOMDV due to the increasing of RREQ and RREP routing packet sizes. The overall average NRL is increased by 3.2% for SSD-AOMDV compared to SD-AOMDV. The increasing in NRL is negligible compared to the overall performance improvement.

.

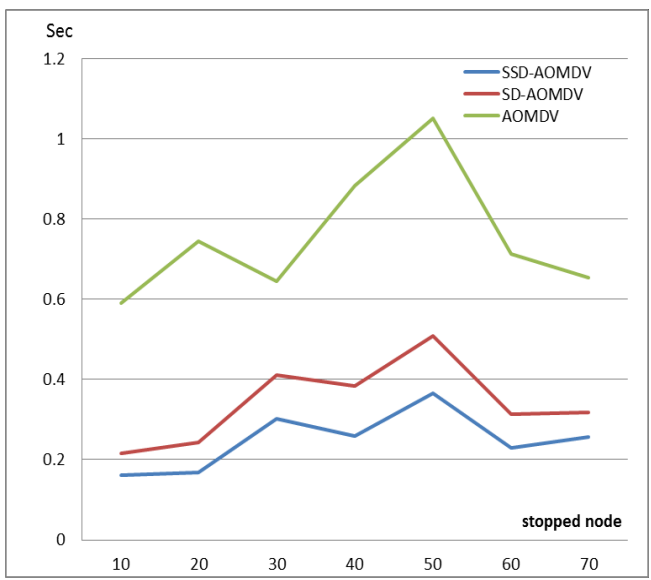

**Fig. 2: End-to-End delay**

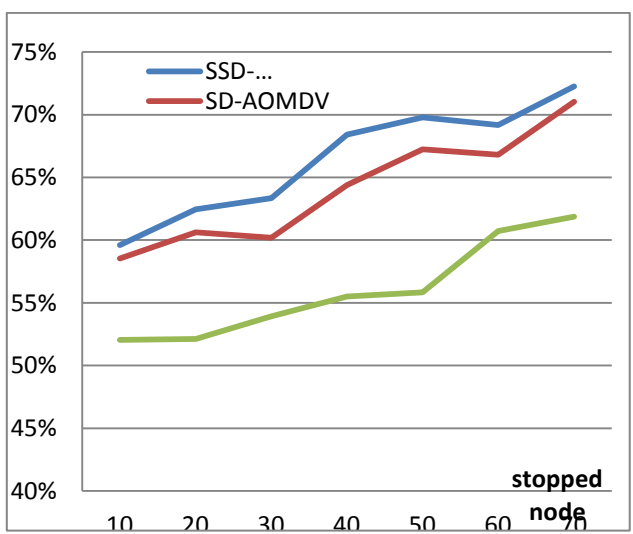

**Fig. 3: PDF**

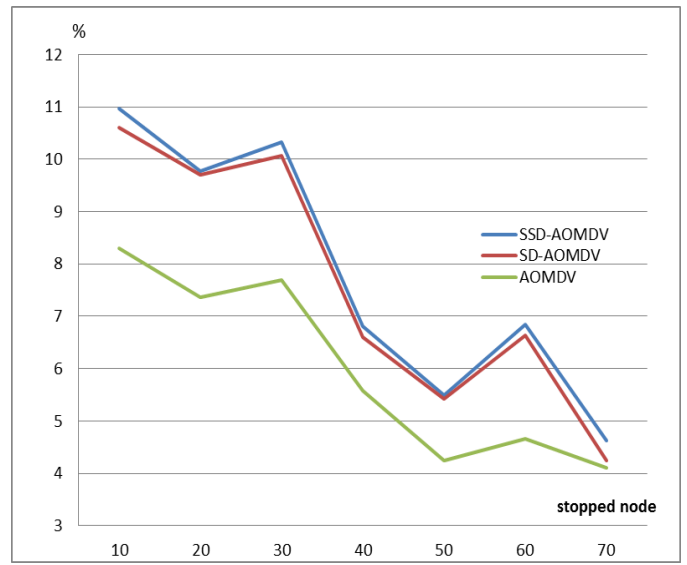

**Fig. 4: NRL**

## 6. CONCLUSIONS

This paper proposes SSD-AOMDV as VANET routing protocol. SSD-AOMDV improves the most important on demand multipath routing protocol AOMDV to suit VANET characteristics. SSD-AOMDV adds the mobility parameters: speed, direction and stop_times to hop count as new AOMDV routing metric to select next hop during the route discovery phase. SSD-AOMDV is designed, implemented, and compared with AOMDV and SD-AOMDV. Simulation results show that SSD-AOMDV has outperformed AOMDV in different traffic scenarios with different percentages of periodically stopped nodes that simulate buses mobility pattern. In our future work, mobile nodes will periodically investigate the traffic environment. Based on the road traffic conditions, nodes will configure the most appropriate routing protocols AOMDV, SD-AOMDV, or SSD-AOMDV to suit the current traffic pattern.

## 7. REFERENCES

[1] Ramakrishnan , R. S. Rajesh , R. S. Shaji,"CBVANETA Cluster Based Vehicular Adhoc Network Model for Simple Highway Communication" J. Advanced Networking and Applications Volume: 02, Issue: 04, Pages: 755-761 (2011).

[2] Yun-Wei Lin1, Yuh-Shyan Chen, and Sing-Ling Lee1," Routing Protocols in Vehicular Ad Hoc Networks: A Survey and Future Perspectives "Journal of Information Science And Engineering 26, 913-932 (2010).

[3] .J. Lou, J.P. Hubaux, " A Survey of Inter-Vehicle Communication", Technical Report, School of Computer and Communication Science, EPFL, Switzerland, 2004.

[4] D. Johnson et al., "Dynamic Source Routing for Mobile Ad Hoc Networks", IEFT MANET Draft, April 2003.

[5] Royer et al., "A review of current routing protocols for ad hoc mobile wireless networks", IEEE Personal Communications, Apr'99.

[6] C. E. Perkins and E. M. Royer, " Ad-hoc On-Demand Distance Vector Routing", In Proceedings of the 2nd IEEE Workshop on Mobile Computing Systems and Applications, pages 90–100, New Orleans, LA, 1999.

[7] C. Perkins, "Ad Hoc On Demand Distance Vector (AODV) routing", Internet-Draft, draft – ietf - MANET - aodv-00. Txt, 1997.

[8] C. Perkins, E. Royer, and S. Das., "Ad hoc on-demand distance vector (AODV) routing". Internet Draft, Internet Engineering Task Force, Mar. 2001.

[9] M. K. Marina, and S. R. Das, "On-demand multipath distance vector routing in Ad Hoc networks," Proc. 9th International Conference on Network Protocols, IEEE Press, Nov. 2001, pp. 14-23, doi: 10.1109/ICNP.2001.992756.

[10] R. Biradar, Koushik Majumder, Subir Kumar Sarkar, Puttamadappa, "Performance Evaluation and Comparison of AODV and AOMDV" S.R.Biradar et al. / (IJCSE) International Journal on Computer Science and Engineering Vol. 02, No. 02, 2010, 373-377.

[11] Hafez Moawad; Eman Shaaban "Efficient Routing Protocol for Vehicular Ad Hoc Networks " 2012 9th IEEE International Conference on Networking, Sensing and Control, China.

[12] Naumov, Valery, and Thomas R. Gross. "Connectivity-aware routing (CAR) in vehicular ad-hoc networks." INFOCOM 2007. 26th IEEE International Conference on Computer Communications. IEEE. IEEE, 2007.

[13] Lin, Yuh-Chung, and Chu-Wei Ke. "Adaptive Route Selection in mobile ad hoc networks." Fourth International Conference on Communications and Networking, pp. 1-5. IEEE, China, 2009.

[14] Shin, Ducksoo, Jonghyup Lee, Jaesung Kim, and JooSeok Song. "A2OMDV: An Adaptive Ad hoc On-demand Multipath Distance Vector Routing Protocol Using Dynamic Route Switching." Executive Development 21 (2008): 22.

[15] Abedi. O., Fathy M., Taghiloo J. "Enhancing AODV Routing Protocol Using Mobility Parameters in VANET" Computer Systems and Applications, 2008. AICCSA 2008. IEEE/ACS International Conference.

[16] http://www.isledu/nsnam/ns.

[17] http://vanet.eurecom.fr.